\documentstyle[emulateapj,psfig]{article}
\received{4th August 1997}
\revised{17th August 1997}
\accepted{---}
\articleid{}{}
\slugcomment{\hbox{\sc Accepted Version}}
\def\gs{\mathrel{\raise0.35ex\hbox{$\scriptstyle >$}\kern-0.6em 
\lower0.40ex\hbox{{$\scriptstyle \sim$}}}}
\def\ls{\mathrel{\raise0.35ex\hbox{$\scriptstyle <$}\kern-0.6em 
\lower0.40ex\hbox{{$\scriptstyle \sim$}}}}
\def\et{\hbox{et al.}$\,$}
\lefthead{Smail, Ivison \& Blain}
\righthead{A Deep Sub-millimeter Survey of Lensing Clusters}
\begin{document}

\title{A Deep Sub-millimeter Survey of Lensing Clusters:\\  
A New Window on Galaxy Formation and Evolution}

\author{Ian Smail,$\!$\altaffilmark{1,2} R.\,J.\ Ivison\altaffilmark{2,3}
\& A.\,W.\ Blain\altaffilmark{4}}
\affil{\tiny 1) Department of Physics, University of Durham, South Rd, Durham DH1 3LE, UK}
\affil{\tiny 3) Institute for Astronomy, Dept.\ of Physics \& Astronomy, 
       University of Edinburgh, Blackford Hill, Edinburgh EH9 3HJ, UK}
\affil{\tiny 4) Cavendish Laboratory, Madingley Rd, Cambridge
CB3 OHE, UK}

\altaffiltext{2}{PPARC Advanced Fellow.}
\setcounter{footnote}{4}

\begin{abstract}
We present the first results of a sub-millimeter survey of distant
clusters using the new Sub-mm Common-User Bolometer Array (SCUBA) on
the James Clerk Maxwell Telescope.  We have mapped fields in two
massive, concentrated clusters, A370 at $z=0.37$ and Cl\,2244$-$02 at
$z=0.33$, at wavelengths of 450 and 850\,$\mu$m.  The resulting
continuum maps cover a total area of about 10\,arcmin$^2$ to 1$\sigma$
noise levels less than 14 and 2\,mJy\,beam$^{-1}$ at the two
wavelengths, 2--3 orders of magnitude deeper than was previously
possible. We have concentrated on lensing clusters to exploit the
amplification of {\it all} background sources by the cluster, improving
the sensitivity by a factor of 1.3--2 as compared with a blank-field
survey.  A cumulative source surface density of $(2.4\pm 1.0) \times
10^3$ degree$^{-2}$ is found to a 50\% completeness limit of $\sim
4$\,mJy at 850\,$\mu$m.  The sub-mm spectral properties of these
sources indicate that the majority lie at high redshift, $z>1$.
Without correcting for lens amplification, our observations limit the
blank-field counts at this depth.  The surface density is 3 orders of
magnitude greater than the expectation of a non-evolving model using
the local {\it IRAS} 60\,$\mu$m luminosity function.  The observed
source counts thus require a substantial increase in the number density
of strongly star-forming galaxies in the high-redshift Universe and
suggest that optical surveys  may have substantial underestimated the
star formation density in the distant Universe.  Deeper sub-mm surveys
with SCUBA should detect large numbers of star-forming galaxies at high
redshift, and so provide strong constraints on the formation of normal
galaxies.
\end{abstract}



\section{Introduction}

Surveys of the local Universe have shown that a third of the total
bolometric luminosity is emitted at sub-mm and far-infrared wavelengths
as a result of reprocessing of star-light by dust
(Soifer \& Neugebauer 1991). Moreover, some of the most vigorous
star-forming galaxies in the local Universe are also those in which the
effects of dust obscuration are most significant.  While there have
been striking advances in the identification of `normal' galaxies at
high redshift ($z\sim2$--4) using Lyman-dropout techniques (Steidel \et
1996), such approaches are insensitive to highly obscured star-forming
galaxies at these epochs.   The presence of at least modest amounts of
dust in distant proto-galaxies, especially forming spheroids, is
expected given the highly metal-enriched ISM which must be present
during their formation (e.g.\ Mazzei \& de Zotti 1996). Thus 
techniques sensititive to obscured distant galaxies should be used
to investigate the formation of these populations in the early Universe.

Sensitive sub-mm observations present the first opportunity to detect
dusty star-forming galaxies at high redshift.  At wavelengths around
100\,$\mu$m, the bulk of the luminosity of normal, star-forming
galaxies is reprocessed star-light from dust and so observations in the
sub-mm band can provide robust estimates of both the dust mass and
total star formation rate in a galaxy.  Furthermore, the negative {\it
K}-correction at wavelengths $\lambda \gs 400$\,$\mu$m means that
sub-mm observations select star-forming galaxies at $z\gs 1$ in an
almost distance-independent manner, providing an efficient method for
the detection of star-forming galaxies at very large redshifts, $z \ls
10$, and hence studying galaxy evolution (Blain
\& Longair 1993, 1996 --- BL96; Blain 1997; Eales \& Edmunds 1997; 
Franceschini \et 1997; Guiderdoni \et 1997).

%
%
\begin{table*}
{\scriptsize
\begin{center}
\centerline{Table 1}
\vspace{0.1cm}
\centerline{SCUBA observations of A370 and Cl\,2244$-$02}
\vspace{0.3cm}
\begin{tabular}{lccccccc}
\hline\hline
\noalign{\smallskip}
 {Target} & {R.A.} & {Dec.} & {$\lambda$} &
  {Exposure time} & {Area} &
                      {Flux density} \cr
~ & {(J2000)} & {(J2000)} &  
{($\mu$m)} & {(ks)} & {(arcmin$^2$)} & {1$\sigma$ (mJy)} \cr
\hline
\noalign{\smallskip}
Cl\,2244$-$02 & 22~47~11.9 & $-$02~05~38 & 450 &  23.0
& 4.00 & 14.0 \cr
~ & ~ & ~ & 850 & 23.0 & 5.40 & 1.9 \cr
\noalign{\smallskip}
A370 & 02~39~53.0 & $-$01~35~06 & 450 &  25.7
& 4.35 & 13.3 \cr
~ & ~ & ~ & 850 &  25.7 & 5.40 & 1.8 \cr
\noalign{\smallskip}
\noalign{\hrule}
\noalign{\smallskip}
\end{tabular}
\end{center}
}
\vspace*{-0.8cm}
\end{table*}

Most published sub-mm studies of distant galaxies have targeted
atypical galaxies (e.g.\ radio-loud galaxies, Ivison \et 1998). We
report here the first deep sub-mm survey to probe the nature of normal
galaxies at moderate and high redshift, $z\gs 0.5$.  In this study we
have attempted to maximise the available sample of distant galaxies by
concentrating on fields in moderate-redshift clusters.  While the
dominant spheroidal populations of these clusters are expected to be
quiescent in the sub-mm band, the in-fall of field galaxies associated
with the growth of the clusters (Smail \et 1997) means that these
fields may contain over-densities of moderate-redshift star-forming
galaxies, as compared with `blank' field surveys.  

The main attraction of the clusters observed here, however, is that
they are strong gravitational lenses, magnifying any source
lying behind them (Blain 1997).  Given the expected steep rise in the
sub-mm counts (BL96), amplification bias could increase the source
counts in these fields by a substantial factor, with a maximum
predicted surface density of about 10 sources per SCUBA field down to
1\,mJy at 850\,$\mu$m (Blain 1997).  Moreover, by targeting those
clusters that contain giant arcs, images of distant field galaxies
magnified by factors of 10--20, we can also obtain otherwise
unachievable sensitivity ($\ls 0.1$\,mJy at 850\,$\mu$m) on the dust
properties of a few serendipitously-positioned normal galaxies at high
redshift.  The angular scales of the region where these highly
magnified high-redshift galaxies are found is also well-matched to the
SCUBA field-of-view.

In the following sections we give details of the observations and their
reduction, and discuss the results within the framework of current
theoretical models of galaxy formation and evolution.  We adopt
$H_\circ=50 $\,km s$^{-1}$ Mpc$^{-1}$ and $q_\circ = 0.5$.

\section{Observations and Reduction}

These data were obtained using SCUBA (Cunningham \& Gear 1994) on the
James Clerk Maxwell Telescope (JCMT)\footnote{The JCMT is operated by
the Observatories on behalf of the UK Particle Physics and Astronomy
Research Council, the Netherlands Organization for Scientific Research
and the Canadian National Research Council.}. SCUBA contains a number
of detectors and detector arrays cooled to 0.1\,K and covering the
atmospheric windows from 350\,$\mu$m to 2000\,$\mu$m.   In our survey,
we operated the 91 element Short-wave (SW) array at 450\,$\mu$m and the
37 element Long-wave (LW) array at 850\,$\mu$m, giving half-power beam
widths of 7.5 and 14.7\,arcsec respectively.  Both arrays have a
2.3\,arcmin instantaneous field-of-view and the design of the optics
ensures that, with a suitable jiggle pattern for the secondary mirror,,
fully sampled maps can be obtained simultaneously at 450 and
850\,$\mu$m.  The multiplexing and high efficiency of the arrays means
that SCUBA offers a gain in mapping speed of a factor of about 300 as
compared with previous detectors.

The observations employed a 64-point jiggle pattern, fully sampling
both arrays over a period of 128\,s.  The pattern was subdivided so
that the target position could be switched between the signal and
reference every 32\,s: a repeating signal--reference--reference--signal
scheme, with sixteen 1\,s jiggles in each. Whilst jiggling, the
secondary was chopped at 6.944\,Hz by 60\,arcsec in azimuth.  The
pointing stability was checked every hour and regular skydips were
performed to measure the atmospheric opacity.  The rms pointing errors
were below 2\,arcsec, while atmospheric zenith opacities at 450 and
850\,$\mu$m were very stable during the course of each night, the night
to night variations being in the range 0.98--2.2 and 0.18--0.37
respectively.

The dedicated SCUBA data reduction software (SURF, Jenness 1997) was
used to reduce the observations.  The reduction consisted of
subtracting the reference from the signal after carefully rejecting
spikes and data from noisy bolometers.  Six quiet bolometers at the
edge of each array were used to compensate for spatially-correlated sky
emission. This reduced the effective noise-equivalent flux density from
100--350 to 90\,mJy\,Hz$^{-1/2}$ at 850\,$\mu$m (c.f.\ Ivison \et
1998).  The resulting maps were flatfielded, corrected for atmospheric
attenuation, and calibrated using nightly beam maps of Uranus.  The
calibrated maps from each night were coadded and then linearly
interpolated onto an astrometric grid using an approximately Nyquist
sampling, of 2 and 4\,arcsec\,pixel$^{-1}$ at 450 and 850\,$\mu$m
respectively, to produce the maps presented in Fig.~1 (Plate~1).

The final on-source integration times are listed in Table~1, along with
field positions and sensitivity limits. The beams have moderate error
lobes, especially at 450\,$\mu$m, and so we correct our aperture
measurements for flux outside the aperture using the measured values
from our calibration sources.  Even without including a factor to
account for the lensing amplification, the data shown in Fig.~1 are the
deepest sub-mm maps ever published, and illustrate the cosmetically
clean and flat maps achievable with SCUBA in long integrations.

\section{Analysis and Results}

%
%
\begin{table*}
{\scriptsize
\begin{center}
\centerline{\sc Table 2}
\vspace{0.1cm}
\centerline{\sc Sub-mm source counts}
\vspace{0.3cm}
\begin{tabular}{lccccccc}
\hline\hline
\noalign{\smallskip}
 {Target} & {$\lambda$} &
{Threshold} & {Min.\ Area}   &  {N} & {N$_{-ve}$} &
S$_{80\%}$ & S$_{50\%}$ \cr
~ &  {($\mu$m)}  & {(mJy beam$^{-1}$)} & {(arcsec$^2$)} & ~ & ~ &  {(mJy)} & {(mJy)} \cr
\hline
\noalign{\smallskip}
Cl\, 2244$-$02 & 450 & 16.0 & 16 & 0 & 0 & 135 &  63 \cr
             ~ & 850 & ~1.9 & 64 & 2 & 0 & 5.5 & 3.9 \cr
\noalign{\smallskip}
A370           & 450 & 16.0 & 16 & 1 & 0 & 140 &  84 \cr
             ~ & 850 & ~1.8 & 64 & 4 & 0 & 5.4 & 3.9 \cr
\noalign{\smallskip}
\noalign{\hrule}
\noalign{\smallskip}
\end{tabular}
\end{center}
}
\vspace*{-0.8cm}
\end{table*}

Source catalogs from our fields were constructed using the Sextractor
package (Bertin \& Arnouts 1996). The detection algorithm requires that
the surface brightness in 4 contiguous pixels exceeds a threshold
(chosen as $\sim 1\sigma$ of the sky noise, Table~2), after subtracting
a smooth background signal and convolving the map with a $4\times 4$
pixel top-hat filter.  The numbers of objects (N) detected in each
field are given in Table~2, indicating that our observations are far
from being limited by confusion: there are $\sim 60$ beams per source. 

To assess the contribution of noise to our catalogs we re-ran the
detection algorithm on the negative fluctuations in the map. This gave
a simple estimate of the number of false--positive detections that may
arise from the noise, assuming that the noise properties of the map are
Gaussian.  We estimate that there are no false detections in our
catalogs (N$_{-ve}$, Table~2), and so all the detections are real.  The
presence of the brightest source in the reference beams (60\,arcsec to
the East and West in the 850\,$\mu$m map of A370) was disregarded. This
detection does confirm, however, the reality of positive features at
this faint level, while the absence of any other negative detections
limits the number of luminous sources which can lie in the regions
covered by the reference beams.

Secondly, to determine the completeness of our sample we added a
faint source to the maps repeatedly, re-ran our detection
algorithm and estimated the efficiency of detecting this source as a
function of its flux density. This provides a reliable estimate
of the visibility of a faint compact source in the maps. The template
source was a scaled version of our calibration source, Uranus. The
estimated 80\% and 50\% completeness limits of the catalogs derived
from these simulations are listed in Table~2 as S$_{80\%}$ and
S$_{50\%}$. The incompleteness limits are relatively bright for the
450\,$\mu$m maps because a large proportion of the flux density (about
40\%) is found in the low surface brightness wings of the beam.  
The simulations also indicate that the measured 850\,$\mu$m
flux densities are unbiased and are typically accurate to 10\% at
25\,mJy and 30\% at 4\,mJy.

We discuss the detailed properties of the sources in another paper,
where we also place limits on the dust masses of the numerous
strongly-lensed distant galaxies covered by our maps.  However, we note
that, based on their weak or non-detection at 450\,$\mu$m, all of the
850\,$\mu$m sources in our sample appear to have the sub-mm spectral
characteristics of distant ($z\gs 1$) star-forming galaxies and are
thus unlikely to be associated with the clusters.  

%
%
\hbox{~}
\centerline{\psfig{file=f2.ps,angle=-90,width=3.3in}}
\noindent{\scriptsize
\addtolength{\baselineskip}{-3pt} 
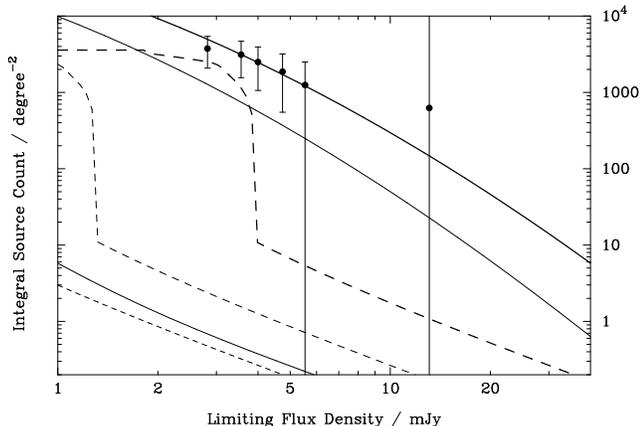
{\bf Figure~2} Models of the integral number counts of
sources at 850\,$\mu$m in a parametric model of galaxy evolution (BL96)
and from a simple model based on the limits on strongly star-forming
systems in optical surveys of distant galaxies.  The observations are
represented by filled circles with Poisson errors, note that the errors
are not independent on the various points.  The observations have been
corrected for the effects of lens amplification using simple lens
models, but not corrected for incompleteness.  The solid curves
represent, in order of increasing predicted counts,  models that
include:  no evolution; $(1+z)^3$ evolution with $z_{\rm max} =$~2 and
$z_0 =$~5 (Model A); and $(1+z)^3$ evolution with $z_{\rm max} =$~2.6
and $z_0 = $~7 (Model B).  The dashed lines represent models where we
fill the Universe across $z=$0--10 with a constant density
(0.6\,$\times$\,10$^{-4}$ Mpc$^{-3}$) of star-forming galaxies with
fixed star formation rates.  In order of increasing predicted counts
the dashed lines represent star formation rates for the population of:
$\stackrel{\bf .}{\rm M} = $~20, 50 and 150~${\rm M}_\odot {\rm
yr}^{-1}$, where we have assumed a dust temperature of $T=60$\,K.
Clearly only models including high densities of strongly star-forming
galaxies are compatible with the observed surface density of sources.

\addtolength{\baselineskip}{3pt}
}

Converting the observed number of sources into a surface density and
correcting for incompleteness, we determine a cumulative number density
across our two fields of $(2.4\pm 1.0)\times 10^3$ degree$^{-2}$ down
to a 50\% completeness limit of 4\,mJy at 850\,$\mu$m (all errors
include only Poisson contributions).  At 450\,$\mu$m, the single
source we detect places only a very weak limit on the likely surface density
of the order of $1 \times 10^3$ degree$^{-2}$ brighter than 80\,mJy.  

We now estimate the likely lens amplification factors, and so place
tighter limits on the typical blank-field counts.  Because the
distances to the detected sources are unknown, this estimate will, by
necessity, be crude and so we have not attempted a detailed analysis.
The cluster potentials are modelled as isothermal spheres with masses
and centers determined from the redshifts and observed shapes of the
giant arc in each cluster (Kneib \et 1993; Smail \et 1996).  In these
models the mean amplification factors for typical background sources
($z\gs 1$) are about 2 and 1.3 in the regions covered by our maps of
A370 and Cl\,2244$-$02 respectively, while the observed area of the
maps (5.4 arcmin$^2$ at 850\,$\mu$m) corresponds to 1.8 and
4.0\,arcmin$^2$ in the respective source planes.  Correcting the flux
densities of our sources to take account of the probable lens
amplifications, but not correcting for incompleteness, we predict the
source counts presented in Fig.~2.  These indicate integrated number
densities in blank fields of $(2.5 \pm 1.4) \times 10^3$ (similar to
the observed value due to the source counts having a form close to
$S^{0.4}$) and $(3.6\pm 1.6) \times 10^3$ degree$^{-2}$ to flux limits
of 4 and 3\,mJy respectively at 850\,$\mu$m.

~From the flux densities associated with the resolved sources in the
fields we calculate lower limits to the background radiation
intensities of  2.6 and $2.4 \times 10^{-10}$\,W\,m$^{-2}$\,sr$^{-1}$
at wavelengths of 450\,$\mu$m and 850\,$\mu$m respectively, averaged
over both fields.  By extrapolating the 850\,$\mu$m counts using our
best-fit model (Fig.~2 and \S 4) to faint flux densities, we estimate
total intensities of extragalactic background radiation from discrete
sources of about 26--28 and 4.4--$6.7 \times
10^{-10}$\,W\,m$^{-2}$\,sr$^{-1}$ at wavelengths of 450 and 850\,$\mu$m
respectively. These background radiation intensities are broadly
consistent with the tentative detection of an isotropic component of
the background radiation in the sub-mm by Puget \et (1996), who
inferred $\nu I_\nu = (23 \pm 20)$ and $(2.7 \pm 2.0) \times
10^{-10}$\,W\,m$^{-2}$\,sr$^{-1}$ at wavelengths of 450 and 850\,$\mu$m
respectively.  If we assume that all of the background radiation
intensity we infer is due to the formation of massive stars, then we
expect that a density parameter of heavy elements of $\ls 6 \times
10^{-4}$ will have accumulated in the Universe by the present epoch.
This density corresponds to about 1.1\% of the density parameter in
baryons if $\Omega_{\rm b} = 0.05$, and so it is fully consistent with
present limits.  We reiterate, however, that these are tentative
estimates, the accuracy of which depends on the models assumed for both
the lens and the form of the counts of distant galaxies.

\section{Discussion}

Due to the negative {\it K}-corrections expected for distant galaxies,
sub-mm observations provide a good estimate of the volume density of
luminous star-forming galaxies at $z\gs 1$.  In the absence of
redshifts for all the sources, the evolution of this population can be
understood by comparing parameterised models to the counts (BL96). The
BL96 models are based on the 60\,$\mu$m luminosity function of {\it
IRAS} galaxies (Saunders \et 1990) and assume that the luminosities
evolve as $(1+z)^3$ out to a redshift, $z_{\rm max}$, and then maintain
the enhanced luminosity out to a cutoff redshift, $z_0$.   The form of
this evolution is motivated by the observations of similar behaviour in
both the radio galaxy and QSO number counts (Dunlop \& Peacock 1991) as
well as the luminosity density of field galaxies at $z<1$ (Lilly \et
1996).  BL96 also give predicted counts for a non-evolving model using
the same luminosity function.  The adopted parameters for the models
described in that paper give source counts which roughly span the range
predicted by other similar works (e.g.\ Guiderdoni \et 1997).  In
Fig.~2, we plot both the no-evolution case and two other parametric
models based on BL96: model A -- Model~2 in BL96 -- with values of
$z_{\rm max}=2$ and $z_0=5$; and model B, has $z_{\rm max}\simeq 2.6$
and $z_0=7$, although most combinations of $z_{\rm max}\simeq 2.2$--2.9
and $z_0 \gs 5$ give comparable results.  Model B was used to estimate
the total background radiation intensity discussed in \S3.

~From Fig.~2 it can be seen that the no evolution predictions fall
short by 2--3 orders of magnitude of the observations.  Thus, this
first analysis of a deep sub-mm survey indicates that the number
density of strongly star-forming galaxies, and so the mean
star formation rate in the distant Universe is considerably larger than
that seen locally.  To estimate the extent of this evolution we assume
that all the detected sources lie beyond the clusters.   We then
require strong evolution, of the form given in model B, out to $z>2$ to
fit the 850\,$\mu$m counts.  For consistency, we check the
predictions from model B for the observed counts at 450\,$\mu$m; 
0.7 sources are expected in the two fields, in agreement with the
single detection.

We conclude from the 850\,$\mu$m counts that the integrated star
formation rate in the Universe, as traced by the number density of the
most luminous sources, must continue to rise out to beyond $z> 1$ ($z>
2$ in our parametric models), extending the trend observed at $z<1$
(Lilly \et 1996).  For sources at $z\gs 1$, the typical luminosity we
infer is $L_{\rm FIR} \sim 0.5$--$1.0 \times 10^{13} {\rm L}_\odot$,
with a star formation rate of $\stackrel{\bf .}{\rm M} \gs 100$--$300
{\rm M}_\odot$\,yr$^{-1}$.  Using the observed surface density of these
objects and assuming a constant space density of sources between $z=
1$--5, we would predict a number density of strongly star-forming
galaxies of:  $N(\stackrel{\bf
.}{\rm M} \gs 150 {\rm M}_\odot {\rm yr}^{-1}) \sim 1.2 \times 10^{-4}$
Mpc$^{-3}$, at $z\gs 1$.  

Limits on the number density of strongly star-forming galaxies at
$z\sim 2$--3.5 have recently been published by Madau \et~(1996 --- M96)
on the basis of Lyman-dropout surveys.  Their limit is $N(\stackrel{\bf
.}{\rm M} > 20 {\rm M}_\odot {\rm yr}^{-1}) < 0.6 \times 10^{-4}$
Mpc$^{-3}$.  We plot in Fig.~2 three models using this number density
of sources to uniformly populate the volume from $z=0$--10, but
allowing the corresponding star formation rate to vary:  $\stackrel{\bf
.}{\rm M} = 20$, 50 and $150 {\rm M}_\odot {\rm yr}^{-1}$. These
values represent that originally given by M96, the M96 limit corrected
for dust extinction as suggested by Pettini \et~(1997), and a mean star
formation rate closer to that needed to fit our observations.  A galaxy
population consistent with the limits from Madau \et~predicts a source
density 3 orders of magnitude lower than that observed (Fig.~2).  Even
if the modest dust extinction proposed by Pettini \et~is included, we still
under-predict the observed surface densities (unless the dust in this
population is very cold, $T=40$\,K, and they have very large dust
masses).  To match the observed surface density of 850\,$\mu$m sources
we must significantly increase the mean star formation, either by
further increasing the star formation rate associated with the
optically-selected samples (c.f.\ Meurer \et 1997) or by introducing a
population of strongly star-forming, but highly obscured, distant
galaxies missed in these samples.   Deeper sub-mm surveys (BL96;
Pearson \& Rowan-Robinson 1996) are necessary to differentiate between
these possibilities and hence provide an unbiased view of star
formation in the distant Universe.

\section{Conclusions}

\noindent{$\bullet$} We have presented the first sub-mm survey of the
distant Universe, deep enough that we should detect the evolving
galaxies predicted by current theoretical models, while at the same
time covering a sufficiently large area to be statistically reliable.
We derive cumulative source counts of $(2.4\pm 1.0) \times 10^3$
degree$^{-2}$ down to 4\,mJy at 850\,$\mu$m.

\noindent{$\bullet$} The surface density of faint sources in the sub-mm
far exceeds a simple non-evolving model using the locally observed
60\,$\mu$m galaxy luminosity function.  Thus our observations require a
substantial increase in the number density of strongly star-forming
galaxies at $z\gs 1$.  

\noindent{$\bullet$} Comparison of our observations with the
predictions of simple parametric models implies that the luminosity
density of the brightest sub-mm sources continues to increase out to
$z> 1$.  Models based upon the claimed properties of star-forming
galaxies from optically-selected samples of distant galaxies (Madau \et
1996) significantly underestimate the observed surface density of
sub-mm sources.  We suggest that such samples may be missing a
considerable proportion of the star formation in dust-obscured galaxies
at high redshift.  We conclude that question of the evolution of the
star formation density in the distant Universe and hence the epoch of
galaxy formation is still very much open.

\section*{Acknowledgements}

We thank the SCUBA development and commissioning team for providing an
efficient and user-friendly instrument, and Ian Robson for his support,
enthusiasm and drive to see the best science done with the JCMT and
SCUBA.    We wish to thank Richard Bower, Richard Ellis, Carlos Frenk,
Wayne Holland,
Tim Jenness, Jean-Paul Kneib and Malcolm Longair for useful
conversations and help. We also thank an anonymous referee for 
comments that clarified the content and presentation of this paper.




%
%

\centerline{\psfig{file=f1.ps,width=6.0in}}
\vspace*{0.1cm}

\noindent{\scriptsize \addtolength{\baselineskip}{-3pt}
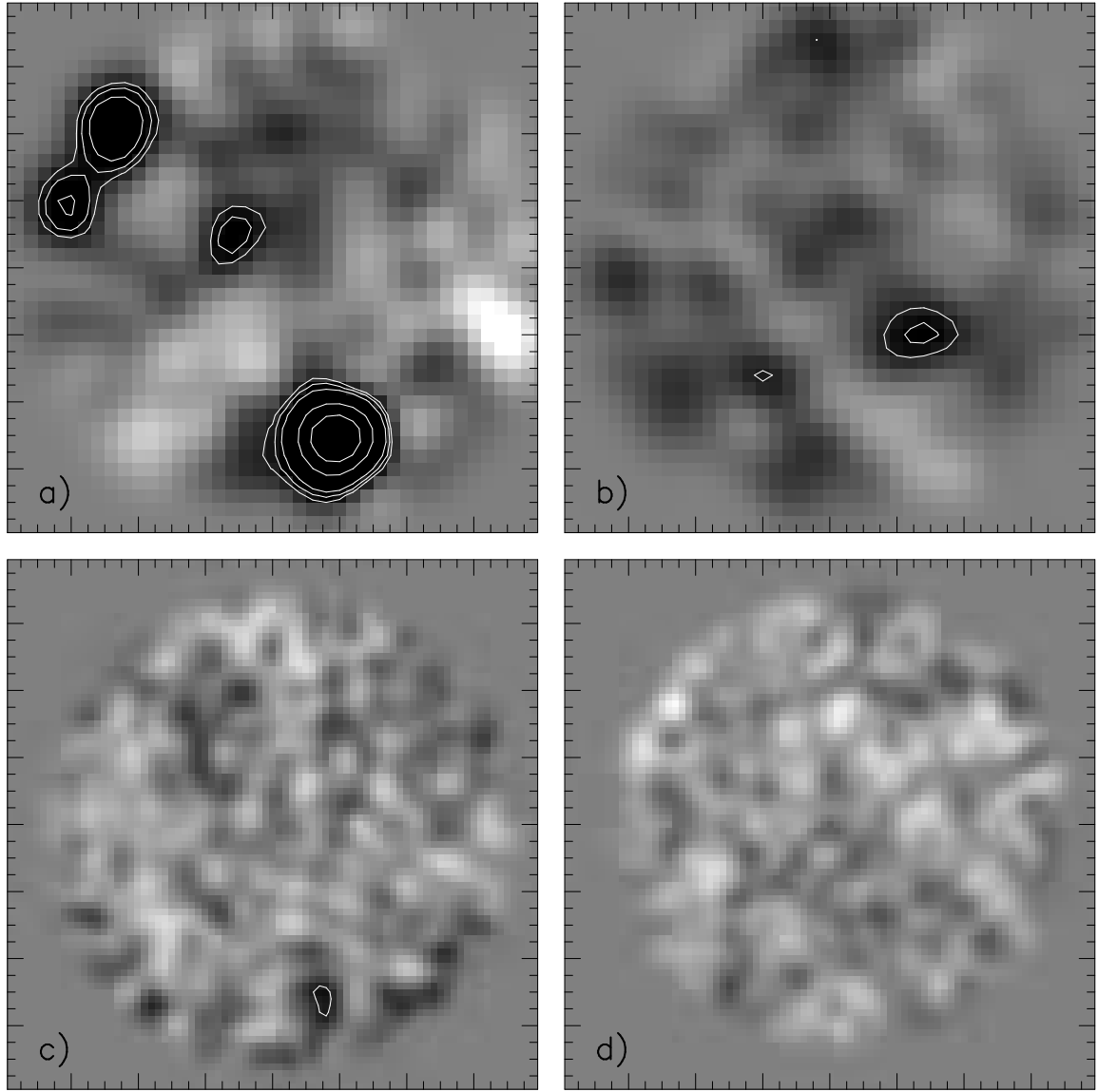
{\bf Figure~1 (Plate 1).} The 450 and 850\,$\mu$m maps of the
two fields: a) A370, 850\,$\mu$m; b) Cl\,2244$-$02, 850\,$\mu$m; c)
A370, 450\,$\mu$m; d) Cl\,2244$-$02, 450\,$\mu$m.  The maps are
smoothed to the instrumental resolution at each wavelength and have had
their boundaries apodized to remove regions on the outskirts with
low effective exposure times resulting from the rotation of the 
field-of-view during our long exposures.   They are
displayed as a grayscale from $-$4$\sigma$ to 4$\sigma$, the contours
are positive and show 3, 4, 5, 10, 15 $\sigma$ for each field.  The
major tick marks are 20 arcsec in all panels.

\addtolength{\baselineskip}{3pt}
}
\bigskip

\end{document}